\documentclass[twocolumn,tighten]{aastex62}
\newcommand{\teff}{$T_{\mathrm{eff}}$}
\newcommand{\muhz}{$\mu$Hz}
\newcommand{\numax}{$\nu_{\mathrm{max}}$}
\newcommand{\dnu}{$\Delta\nu$}
\newcommand{\msol}{M$_\odot$}

\newcommand{\kepler}{\textit{Kepler}}
\newcommand{\keplers}{\textit{Kepler's}}

\received{January 1, 2018}
\revised{January 7, 2018}
\accepted{\today}
\submitjournal{ApJL}

\shorttitle{Oscillating red giants with TESS}
\shortauthors{Silva Aguirre et al.}

\begin{document}

\title{Detection and characterisation of oscillating red giants: first results from the TESS satellite}

\correspondingauthor{V\'ictor Silva Aguirre}
\email{victor@phys.au.dk}

\author[0000-0002-6137-903X]{V\'ictor Silva Aguirre}
\affiliation{Stellar Astrophysics Centre (SAC), Department of Physics and Astronomy, Aarhus University, Ny Munkegade 120, 8000 Aarhus C, Denmark}

\author{Dennis Stello}
\affiliation{School of Physics, The University of New South Wales, Sydney NSW 2052, Australia}
\affiliation{Sydney Institute for Astronomy (SIfA), School of Physics, University of Sydney, NSW 2006, Australia}
\affiliation{Stellar Astrophysics Centre (SAC), Department of Physics and Astronomy, Aarhus University, Ny Munkegade 120, 8000 Aarhus C, Denmark}

\author[0000-0002-5496-365X]{Amalie Stokholm}
\affiliation{Stellar Astrophysics Centre (SAC), Department of Physics and Astronomy, Aarhus University, Ny Munkegade 120, 8000 Aarhus C, Denmark}
\author[0000-0001-9234-430X]{Jakob R. Mosumgaard}
\affiliation{Stellar Astrophysics Centre (SAC), Department of Physics and Astronomy, Aarhus University, Ny Munkegade 120, 8000 Aarhus C, Denmark}

\author[0000-0002-4773-1017]{Warrick H. Ball}
\affiliation{School of Physics and Astronomy, University of Birmingham, Birmingham, B15 2TT, United Kingdom}
\affiliation{Stellar Astrophysics Centre (SAC), Department of Physics and Astronomy, Aarhus University, Ny Munkegade 120, 8000 Aarhus C, Denmark}
\author[0000-0002-6163-3472]{Sarbani Basu}
\affiliation{Department of Astronomy, Yale University, New Have, CT 06520, USA}
\author[0000-0002-9480-8400]{Diego Bossini}
\affiliation{Instituto de Astrof\'{\i}sica e Ci\^{e}ncias do Espa\c{c}o, Universidade do Porto,  Rua das Estrelas, 4150-762 Porto, Portugal}
\author[0000-0003-0142-4000]{Lisa Bugnet}
\affiliation{IRFU, CEA, Universit\'e Paris-Saclay, F-91191 Gif-sur-Yvette, France}
\affiliation{AIM, CEA, CNRS, Universit\'e Paris-Saclay, Universit\'e Paris Diderot, Sorbonne Paris Cit\'e, F-91191 Gif-sur-Yvette, France}
\author[0000-0002-1988-143X]{Derek Buzasi}
\affiliation{Dept. of Chemistry and Physics, Florida Gulf Coast University, 10501 FGCU Blvd. S., Fort Myers, FL 33965 USA}
\author[0000-0002-4588-5389]{Tiago L. Campante}
\affiliation{Instituto de Astrof\'{\i}sica e Ci\^{e}ncias do Espa\c{c}o, Universidade do Porto,  Rua das Estrelas, 4150-762 Porto, Portugal}
\affiliation{Departamento de F\'{\i}sica e Astronomia, Faculdade de Ci\^{e}ncias da Universidade do Porto, Rua do Campo Alegre, s/n, 4169-007 Porto, Portugal}
\author[0000-0003-1001-5137]{Lindsey Carboneau}
\affiliation{Dept. of Chemistry and Physics, Florida Gulf Coast University, 10501 FGCU Blvd. S., Fort Myers, FL 33965 USA}
\author{William J. Chaplin}
\affiliation{School of Physics and Astronomy, University of Birmingham, Birmingham, B15 2TT, United Kingdom}
\affiliation{Stellar Astrophysics Centre (SAC), Department of Physics and Astronomy, Aarhus University, Ny Munkegade 120, 8000 Aarhus C, Denmark}
\author[0000-0001-8835-2075]{Enrico Corsaro}
\affiliation{INAF – Osservatorio Astrofisico di Catania, Via S. Sofia, 78, 95123 Catania, Italy}
\author{Guy R. Davies}
\affiliation{School of Physics and Astronomy, University of Birmingham, Birmingham, B15 2TT, United Kingdom}
\affiliation{Stellar Astrophysics Centre (SAC), Department of Physics and Astronomy, Aarhus University, Ny Munkegade 120, 8000 Aarhus C, Denmark}
\author{Yvonne Elsworth}
\affiliation{School of Physics and Astronomy, University of Birmingham, Birmingham, B15 2TT, United Kingdom}
\affiliation{Stellar Astrophysics Centre (SAC), Department of Physics and Astronomy, Aarhus University, Ny Munkegade 120, 8000 Aarhus C, Denmark}
\author[0000-0002-8854-3776]{Rafael A. Garc\'ia}
\affiliation{IRFU, CEA, Universit\'e Paris-Saclay, F-91191 Gif-sur-Yvette, France}
\affiliation{AIM, CEA, CNRS, Universit\'e Paris-Saclay, Universit\'e Paris Diderot, Sorbonne Paris Cit\'e, F-91191 Gif-sur-Yvette, France}
\author[0000-0001-8330-5464]{Patrick Gaulme}
\affiliation{Max Planck Institute for Solar System Research, Justus-von-Liebig Weg 3, D-37077 G\"ottingen, Germany}
\affiliation{Department of Astronomy, New Mexico State University, P.O. Box 30001, MSC 4500, Las Cruces, NM 88003-8001, USA}
\author[0000-0002-0468-4775]{Oliver J. Hall}
\affiliation{School of Physics and Astronomy, University of Birmingham, Birmingham, B15 2TT, United Kingdom}
\affiliation{Stellar Astrophysics Centre (SAC), Department of Physics and Astronomy, Aarhus University, Ny Munkegade 120, 8000 Aarhus C, Denmark}
\author[0000-0001-8725-4502]{Rasmus Handberg}
\affiliation{Stellar Astrophysics Centre (SAC), Department of Physics and Astronomy, Aarhus University, Ny Munkegade 120, 8000 Aarhus C, Denmark}
\author[0000-0003-2400-6960]{Marc Hon}
\affiliation{School of Physics, The University of New South Wales, Sydney NSW 2052, Australia}
\author[0000-0003-3627-2561]{Thomas Kallinger}
\affiliation{Institut f\"ur Astrophysik, Universit\"at Wien, T\"urkenschanzstrasse 17, 1180 Vienna, Austria}
\author[0000-0001-5219-7894]{Liu Kang}
\affiliation{Department of Astronomy, Beijing Normal University, 100875 Beijing, PR China}
\author[0000-0001-9214-5642]{Mikkel N. Lund}
\affiliation{Stellar Astrophysics Centre (SAC), Department of Physics and Astronomy, Aarhus University, Ny Munkegade 120, 8000 Aarhus C, Denmark}
\author[0000-0002-0129-0316]{Savita Mathur}
\affiliation{Instituto de Astrof\'{\i}sica de Canarias, E-38200 La Laguna, Tenerife, Spain}
\affiliation{Departamento de Astrof\'{\i}sica, Universidad de La Laguna, E-38206 La Laguna, Tenerife, Spain}
\author[0000-0002-8440-1455]{Alexey Mints}
\affiliation{Leibniz-Institut f\"ur Astrophysik Potsdam (AIP), An der Sternwarte 16, 14482 Potsdam, Germany}
\author[0000-0002-7547-1208]{Benoit Mosser}
\affiliation{LESIA, Observatoire de Paris, Universit\'e PSL, CNRS, Sorbonne Universit\'e, Universit\'e de Paris, 92195 Meudon, France}
\author[0000-0002-9424-2339]{Zeynep \c{C}el\.ik Orhan}
\affiliation{Department of Astronomy and Space Sciences, Science Faculty, Ege University, 35100, Bornova, \.Izmir, Turkey}
\author[0000-0002-9414-339X]{Tha\'ise S. Rodrigues}
\affiliation{Osservatorio Astronomico di Padova -- INAF, Vicolo dell'Osservatorio 5, I-35122 Padova, Italy}
\author{Mathieu Vrard}
\affiliation{Instituto de Astrof\'{\i}sica e Ci\^{e}ncias do Espa\c{c}o, Universidade do Porto,  Rua das Estrelas, 4150-762 Porto, Portugal}
\affiliation{Department of Astronomy, The Ohio State University, 140 West 18th Avenue, Columbus OH 43210, USA}
\author[0000-0002-7772-7641]{Mutlu Y{\i}ld{\i}z}
\affiliation{Department of Astronomy and Space Sciences, Science Faculty, Ege University, 35100, Bornova, \.Izmir, Turkey}
\author[0000-0002-7550-7151]{Joel C. Zinn}
\affiliation{School of Physics, The University of New South Wales, Sydney NSW 2052, Australia}
\affiliation{Kavli Institute for Theoretical Physics, University of California, Santa Barbara, CA 93106, USA}
\affiliation{Department of Astronomy, The Ohio State University, 140 West 18th Avenue, Columbus OH 43210, USA}
\author[0000-0001-5759-7790]{S\.ibel  \"Ortel}
\affiliation{Department of Astronomy and Space Sciences, Science Faculty, Ege University, 35100, Bornova, \.Izmir, Turkey}

\author[0000-0003-4745-2242]{Paul G. Beck}
\affiliation{Institute of Physics, Karl-Franzens University of Graz, NAWI Graz, Universit\"atsplatz 5/II, 8010 Graz, Austria}
\affiliation{Instituto de Astrof\'{\i}sica de Canarias, E-38200 La Laguna, Tenerife, Spain}
\affiliation{Departamento de Astrof\'{\i}sica, Universidad de La Laguna, E-38206 La Laguna, Tenerife, Spain}
\author[0000-0002-0656-032X]{Keaton J. Bell}
\affiliation{DIRAC Institute, Department of Astronomy, University of Washington, Seattle, WA 98195-1580, USA}
\affiliation{NSF Astronomy and Astrophysics Postdoctoral Fellow and DIRAC Fellow}
\author[0000-0002-0951-2171]{Zhao Guo}
\affiliation{Center for Exoplanets and Habitable Worlds, Department of Astronomy and Astrophysics, 525 Davey Laboratory, The Pennsylvania State University, University Park, PA16802, USA}
\author[0000-0002-7614-1665]{Chen Jiang}
\affiliation{School of Physics and Astronomy, Sun Yat-Sen University, No. 135, Xingang Xi Road, Guangzhou, 510275, P. R. China}
\author[0000-0002-3322-5279]{James S. Kuszlewicz}
\affiliation{Max Planck Institute for Solar System Research, Justus-von-Liebig Weg 3, D-37077 G\"ottingen, Germany}
\author{Charles A. Kuehn}
\affiliation{Department of Physics and Astronomy, University of Northern Colorado, Greeley, CO 80639, USA}
\author{Tanda Li}
\affiliation{Sydney Institute for Astronomy (SIfA), School of Physics, University of Sydney, NSW 2006, Australia}
\affiliation{Stellar Astrophysics Centre (SAC), Department of Physics and Astronomy, Aarhus University, Ny Munkegade 120, 8000 Aarhus C, Denmark}
\affiliation{School of Physics and Astronomy, University of Birmingham, Birmingham, B15 2TT, United Kingdom}
\author[0000-0002-8661-2571]{Mia S. Lundkvist}
\affiliation{Stellar Astrophysics Centre (SAC), Department of Physics and Astronomy, Aarhus University, Ny Munkegade 120, 8000 Aarhus C, Denmark}
\author{Marc Pinsonneault}
\affiliation{Department of Astronomy, The Ohio State University, 140 West 18th Avenue, Columbus OH 43210, USA}
\author[0000-0002-4818-7885]{Jamie Tayar}
\affiliation{Institute for Astronomy, University of Hawaii, 2680 Woodlawn Drive, Honolulu, Hawaii 96822, USA}
\affiliation{Hubble Fellow}

\author{Margarida S. Cunha}
\affiliation{Instituto de Astrof\'{\i}sica e Ci\^{e}ncias do Espa\c{c}o, Universidade do Porto,  Rua das Estrelas, 4150-762 Porto, Portugal}
\affiliation{School of Physics and Astronomy, University of Birmingham, Birmingham, B15 2TT, United Kingdom}
\author[0000-0002-1463-726X]{Saskia Hekker}
\affiliation{Max Planck Institute for Solar System Research, Justus-von-Liebig Weg 3, D-37077 G\"ottingen, Germany}
\affiliation{Stellar Astrophysics Centre (SAC), Department of Physics and Astronomy, Aarhus University, Ny Munkegade 120, 8000 Aarhus C, Denmark}
\author{Daniel Huber}
\affiliation{Institute for Astronomy, University of Hawaii, 2680 Woodlawn Drive, Honolulu, Hawaii 96822, USA}
\author{Andrea Miglio}
\affiliation{School of Physics and Astronomy, University of Birmingham, Birmingham, B15 2TT, United Kingdom}
\affiliation{Stellar Astrophysics Centre (SAC), Department of Physics and Astronomy, Aarhus University, Ny Munkegade 120, 8000 Aarhus C, Denmark}
\author[0000-0003-0513-8116]{Mario J. P. F. G. Monteiro}
\affiliation{Instituto de Astrof\'{\i}sica e Ci\^{e}ncias do Espa\c{c}o, Universidade do Porto,  Rua das Estrelas, 4150-762 Porto, Portugal}
\affiliation{Departamento de F\'{\i}sica e Astronomia, Faculdade de Ci\^{e}ncias da Universidade do Porto, Rua do Campo Alegre, s/n, 4169-007 Porto, Portugal}
\author[0000-0003-4538-9518]{Ditte Slumstrup}
\affiliation{Stellar Astrophysics Centre (SAC), Department of Physics and Astronomy, Aarhus University, Ny Munkegade 120, 8000 Aarhus C, Denmark}
\affiliation{European Southern Observatory, Alonso de C\'ordova 3107, Vitacura, Santiago, Chile}
\author{Mark L. Winther}
\affiliation{Stellar Astrophysics Centre (SAC), Department of Physics and Astronomy, Aarhus University, Ny Munkegade 120, 8000 Aarhus C, Denmark}

\author{George Angelou}
\affiliation{Max-Planck-Institut f\"{u}r Astrophysics, Karl Schwarzschild Strasse 1, 85748, Garching, Germany}
\author{Othman Benomar}
\affiliation{Solar Science Observatory, NAOJ, Mitaka, Japan}
\affiliation{Center for Space Science, New York University Abu Dhabi, UAE}
\author{Attila B\'odi}
\affiliation{Konkoly Observatory, Research Centre for Astronomy and Earth Sciences, H-1121 Budapest, Konkoly Thege M. \'ut 15-17, Hungary}
\affiliation{MTA CSFK Lend\"ulet Near-Field Cosmology Research Group}
\author{Bruno L. De Moura}
\affiliation{Instituto Federal do R. G. do Norte - IFRN, Brazil}
\author{S\'ebastien Deheuvels}
\affiliation{IRAP, Universit\'e de Toulouse, CNRS, CNES, UPS, (Toulouse), France}
\author[0000-0002-6526-9444]{Aliz Derekas}
\affiliation{ELTE E\"otv\"os Lor\'and University, Gothard Astrophysical Observatory, Szombathely, Hungary}
\affiliation{MTA-ELTE Exoplanet Research Group, 9700 Szombathely, Szent Imre h. u. 112, Hungary}
\affiliation{Konkoly Observatory, Research Centre for Astronomy and Earth Sciences, H-1121 Budapest, Konkoly Thege M. \'ut 15-17, Hungary}
\author[0000-0001-7801-7484]{Maria Pia Di Mauro}
\affiliation{INAF-IAPS, Istituto di Astrofisica e Planetologia Spaziali, Via del Fosso del Cavaliere 100, 00133 Roma, Italy}
\author{Marc-Antoine Dupret}
\affiliation{STAR Institute, University of Liège, 19C Allée du 6 Août, B-4000 Liège, Belgium}
\author{Antonio Jim\'enez }
\affiliation{Instituto de Astrof\'{\i}sica de Canarias, E-38200 La Laguna, Tenerife, Spain}
\affiliation{Departamento de Astrof\'{\i}sica, Universidad de La Laguna, E-38206 La Laguna, Tenerife, Spain}
\author[0000-0002-4834-2144]{Yveline Lebreton}
\affiliation{LESIA, Observatoire de Paris, Universit\'e PSL, CNRS, Sorbonne Universit\'e, Universit\'e de Paris, 92195 Meudon, France}
\affiliation{Univ Rennes, CNRS, IPR (Institut de Physique de Rennes) - UMR 6251, F-35000 Rennes, France}
\author{Jaymie Matthews}
\affiliation{Department of Physics and Astronomy, University of British Columbia, Vancouver, Canada}
\author{Nicolas Nardetto}
\affiliation{Universit\'e C\^ote d'Azur, Observatoire de la C\^ote d'Azur, CNRS, Laboratoire Lagrange, France}
\author[0000-0001-7804-2145]{Jose D. do Nascimento, Jr.}
\affiliation{Universidade Federal do Rio Grande do Norte, UFRN, Departamento de F\'isica, 59078-970, Natal, RN, Brazil}
\affiliation{Harvard-Smithsonian Center for Astrophysics, 60 Garden St., Cambridge, MA 02138, USA}
\author[0000-0002-2157-7146]{Filipe Pereira}
\affiliation{Instituto de Astrof\'{\i}sica e Ci\^{e}ncias do Espa\c{c}o, Universidade do Porto,  Rua das Estrelas, 4150-762 Porto, Portugal}
\affiliation{Departamento de F\'{\i}sica e Astronomia, Faculdade de Ci\^{e}ncias da Universidade do Porto, Rua do Campo Alegre, s/n, 4169-007 Porto, Portugal}
\author[0000-0002-0588-1375]{Luisa F. Rodr\'iguez D\'iaz}
\affiliation{Stellar Astrophysics Centre (SAC), Department of Physics and Astronomy, Aarhus University, Ny Munkegade 120, 8000 Aarhus C, Denmark}
\author{Aldo M. Serenelli}
\affiliation{Instituto de Ciencias del Espacio (ICE, CSIC), Campus UAB, Carrer de Can Magrans, s/n, 08193 Cerdanyola del Valles, Spain}
\affiliation{Institut d'Estudis Espacials de Catalunya (IEEC), Gran Capita 4, E-08034, Barcelona, Spain}
\author[0000-0001-9715-5727]{Emanuele Spitoni}
\affiliation{Stellar Astrophysics Centre (SAC), Department of Physics and Astronomy, Aarhus University, Ny Munkegade 120, 8000 Aarhus C, Denmark}
\author{Edita Stonkut\.{e}}
\affiliation{Institute of Theoretical Physics and Astronomy, Vilnius University, Saul\.{e}tekio al. 3, 10257 Vilnius, Lithuania}
\author{Juan Carlos Su\'arez}
\affiliation{Dept. F\'isica Te\'orca y del Cosmos. Universidad de Granada. 18006 Granada, Spain}
\affiliation{Instituto de Astrofísica\'isica de Andaluc\'ia (CSIC). Glorieta de la Astronom\'ia s/n. 18008. Granada, Spain}
\author{Robert Szab\'o}
\affiliation{Konkoly Observatory, Research Centre for Astronomy and Earth Sciences, H-1121 Budapest, Konkoly Thege M. \'ut 15-17, Hungary}
\affiliation{MTA CSFK Lend\"ulet Near-Field Cosmology Research Group}
\author[0000-0001-5542-8870]{Vincent Van Eylen}
\affiliation{Mullard Space Science Laboratory, University College London, Holmbury St Mary, Dorking RH5 6NT, UK}
\author[0000-0002-5152-0482]{Rita Ventura}
\affiliation{INAF – Osservatorio Astrofisico di Catania, Via S. Sofia, 78, 95123 Catania, Italy}
\author[0000-0003-0970-6440]{Kuldeep Verma}
\affiliation{Stellar Astrophysics Centre (SAC), Department of Physics and Astronomy, Aarhus University, Ny Munkegade 120, 8000 Aarhus C, Denmark}
\author[0000-0002-3843-1653]{Achim Weiss}
\affiliation{Max-Planck-Institut f\"{u}r Astrophysics, Karl Schwarzschild Strasse 1, 85748, Garching, Germany}
\author[0000-0001-6832-4325]{Tao Wu}
\affiliation{Yunnan Observatories, Chinese Academy of Sciences, 396 Yangfangwang, Guandu District, Kunming, 650216, P. R. China}
\affiliation{Key Laboratory for the Structure and Evolution of Celestial Objects, Chinese Academy of Sciences, 396 Yangfangwang, Guandu District, Kunming, 650216, P. R. China}
\affiliation{Center for Astronomical Mega-Science, Chinese Academy of Sciences, 20A Datun Road, Chaoyang District, Beijing, 100012, P. R. China}

\author[0000-0001-7139-2724]{Thomas Barclay}
\affiliation{NASA Goddard Space Flight Center, 8800 Greenbelt Rd, Greenbelt, MD 20771, USA}
\affiliation{University of Maryland, Baltimore County, 1000 Hilltop Cir, Baltimore, MD 21250, USA}
\author{J\o rgen Christensen-Dalsgaard}
\affiliation{Stellar Astrophysics Centre (SAC), Department of Physics and Astronomy, Aarhus University, Ny Munkegade 120, 8000 Aarhus C, Denmark}
\author[0000-0002-4715-9460]{Jon M. Jenkins}
\affiliation{NASA Ames Research Center, Moffett Field, CA, 94035, USA}
\author{Hans Kjeldsen}
\affiliation{Stellar Astrophysics Centre (SAC), Department of Physics and Astronomy, Aarhus University, Ny Munkegade 120, 8000 Aarhus C, Denmark}
\author{George R. Ricker}
\affiliation{Department of Physics and Kavli Institute for Astrophysics and Space Research, Massachusetts Institute of Technology, Cambridge, MA 02139, USA}
\author[0000-0002-6892-6948]{Sara Seager}
\affiliation{Department of Physics and Kavli Institute for Astrophysics and Space Research, Massachusetts Institute of Technology, Cambridge, MA 02139, USA}
\affiliation{Department of Earth, Atmospheric and Planetary Sciences, Massachusetts Institute of Technology, Cambridge, MA 02139, USA}
\affiliation{Department of Aeronautics and Astronautics, MIT, 77 Massachusetts Avenue, Cambridge, MA 02139, USA}
\author[0000-0001-6763-6562]{Roland Vanderspek}
\affiliation{Department of Physics and Kavli Institute for Astrophysics and Space Research, Massachusetts Institute of Technology, Cambridge, MA 02139, USA}



\begin{abstract}
Since the onset of the `space revolution' of high-precision high-cadence photometry, asteroseismology has been  demonstrated as a powerful tool for informing Galactic archaeology investigations. The launch of the NASA TESS mission has enabled seismic-based inferences to go full sky -- providing a clear advantage for large ensemble studies of the different Milky Way components. Here we demonstrate its potential for investigating the Galaxy by carrying out the first asteroseismic ensemble study of red giant stars observed by TESS. We use a sample of 25 stars for which we measure their global asteroseimic observables and estimate their fundamental stellar properties, such as radius, mass, and age.  Significant improvements are seen in the uncertainties of our estimates when combining seismic observables from TESS with astrometric measurements from the Gaia mission compared to when the seismology and astrometry are applied separately. Specifically, when combined we show that stellar radii can be determined to a precision of a few percent, masses to 5-10\% and ages to the 20\% level. This is comparable to the precision typically obtained using end-of-mission \kepler\ data.
\end{abstract}
\section{Introduction} \label{sec:intro}
Asteroseismology of red giant stars has been one of the major successes of the CoRoT 
and \kepler\ missions. The unambiguous detection of non-radial oscillations has fundamentally widened our understanding of the inner workings of red giants, including the conditions in their core \citep[e.g.,][]{2011Natur.471..608B}. The observed frequency spectra have allowed the determination of the physical properties of thousands of red giants to an unprecedented level of precision \citep[e.g.,][]{2013MNRAS.429..423M}, paving the way for the emergence of asteroseismology as a powerful tool for Milky Way studies and Galactic archaeology \citep[e.g.,][]{Miglio09,2016MNRAS.455..987C,Anders17,2018MNRAS.475.5487S,2019arXiv190412444S}. The {\it Transiting Exoplanet Survey Satellite} \citep[TESS,][]{2015JATIS...1a4003R} is on the path of continuing this legacy with its all-sky survey that is expected to increase the number of detected oscillating red giants by an order of magnitude compared to the tens of thousands reported by its predecessors CoRoT and \kepler.

In the nominal TESS mission, the ecliptic northern and southern hemispheres are each observed during thirteen 27-day-long sectors, and most (92\%) of the surveyed sky will be monitored for just 1-2 sectors. Except for the 20,000 targets pre-selected in each sector for 2-min cadence observations, all stars are observed as part of the full frame images obtained in 30-min cadence, similar to the long cadence sampling of the \kepler\ satellite. The length of the observations sets the lower limit on the oscillation frequencies one can resolve, and the sampling sets the upper frequency limit. We know from previous \kepler\ observations that one month of 30-min cadence data should be well suited to detect oscillations in the low red-giant branch and sufficient to measure the global oscillation properties characterising the frequency spectrum, in particular, its frequency of maximum power, \numax, and the frequency separation between overtone modes, \dnu\ \citep{Bedding:2010ki}. These in turn can be used in combination with complementary data such as the effective temperature, \teff, the relative iron abundance, [Fe/H], and parallax, to obtain precise stellar properties (including ages) when applying asteroseismic-based grid modelling approaches \citep[see e.g.,][]{2017MNRAS.467.1433R,2018ApJS..239...32P}.

Due to the large sky coverage, approximately 97\% of asteroseismic detections in red giants from the TESS nominal mission data are expected to come from stars observed for only one or two sectors\footnote{Based on a preliminary simulation of the full TESS sky (TESS GI Proposal No G011188).}. Here we set out to explore the capability of TESS to detect the oscillations in giants ranging from the base of the red giant branch to the red clump, determine their stellar properties, and use that to assess the prospects for Galactic archaeology studies using one to two sectors of TESS data.
\section{Target selection}\label{sec:targets}
\begin{figure}
\includegraphics[width=\linewidth]{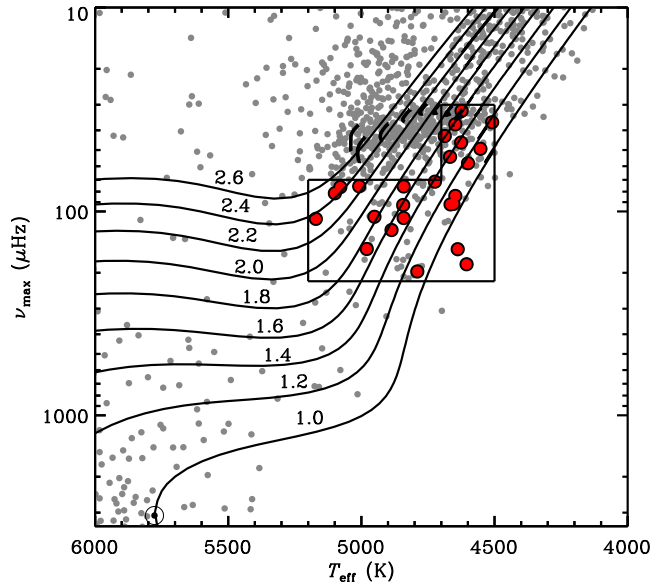}
\caption{`Asteroseismic HR diagram' showing (predicted) \numax\ instead of luminosity. Red dots show the selected targets inside the black selection box. For reference, the Sun is shown as well as all {\it Hipparcos} stars brighter than 6th magnitude (grey dots). Solar metalicity MESA tracks from \citet{Stello:2013jz} are shown to guide the eye with masses in solar units indicated (pre- and post- helium core-ignition phases are shown separately).\label{Fig:hrd}}
\end{figure}
Our goal is to have a representative sample of giants including the types of stars in which we can expect to detect oscillations from one sector 30-min cadence TESS data. We selected red-giant candidates observed during sectors 1 and/or 2 that were deemed viable for asteroseismic detections according to their predicted properties based on the {\it Hipparcos} catalogue \citep{VanLeeuwen:2007dc}. We first estimated the stellar \teff\ and luminosity using $B-V$ color, $V$-band, and {\it Hipparcos} parallax, and the color-temperature and bolometric correction relations of \citet{1996ApJ...469..355F}. We then obtained a prediction of \numax\ ($\propto$ \teff$^{3.5} M/L$; solar scaled, e.g. \citealt{2018ApJS..236...42Y}) for each star assuming a mass of 1.2~\msol, which is representative of a typical red giant as observed by \kepler\ \citep[and unlikely to be more than a factor of two from the true value of each star, e.g.][]{2018ApJS..236...42Y}. We note that one of our targets (TIC~129649472) is a known exoplanet host star recently analysed by \citet{2019arXiv190905961C}.

To ensure that the selected targets were amenable to asteroseismic detection from one sector of 30-min cadence data, we required that they would have an expected \numax\ in the range 30-220\muhz\ and \teff\ in the typical range of red giants of 4500-5200$\,$K. In addition, we applied a narrower \teff\ range of 4500-4700$\,$K for the stars with \numax\ between 30\muhz\ and 70\muhz, to avoid having red clump stars dominating our sample. The resulting sample of stars span evolutionary phases from the base of the red giant branch to the red giant branch bump, as well as some clump stars. 

From this sample, we selected the 25 brightest targets for light curve extraction and asteroseismic analysis. The faintest stars in our sample turned out to be $\sim$6-7th magnitude in $V$ band (see Table~\ref{tab:props}). Under the assumption that the photometric performance of TESS is similar to \keplers, apart from its smaller aperture, this magnitude limit is equivalent to 11-12th magnitude for \kepler. Because single-quarter observations from \keplers\ second life, K2, showed no oscillation detection bias for red giants brighter than around 12th magnitude \citep{2017ApJ...835...83S} we would expect to detect oscillations in all 25 giants with TESS.

Figure~\ref{Fig:hrd} illustrates the location of the selected stars in the HR-diagram and the applied selection criteria. We confirmed that the stars were in sectors 1-2 using the Web TESS Viewing tool (WTV)\footnote{\url{https://heasarc.gsfc.nasa.gov/cgi-bin/tess/webtess/wtv.py}}.
\section{Data processing and asteroseismic analysis} \label{sec:data_analysis}
\begin{figure}
\includegraphics[width=\linewidth]{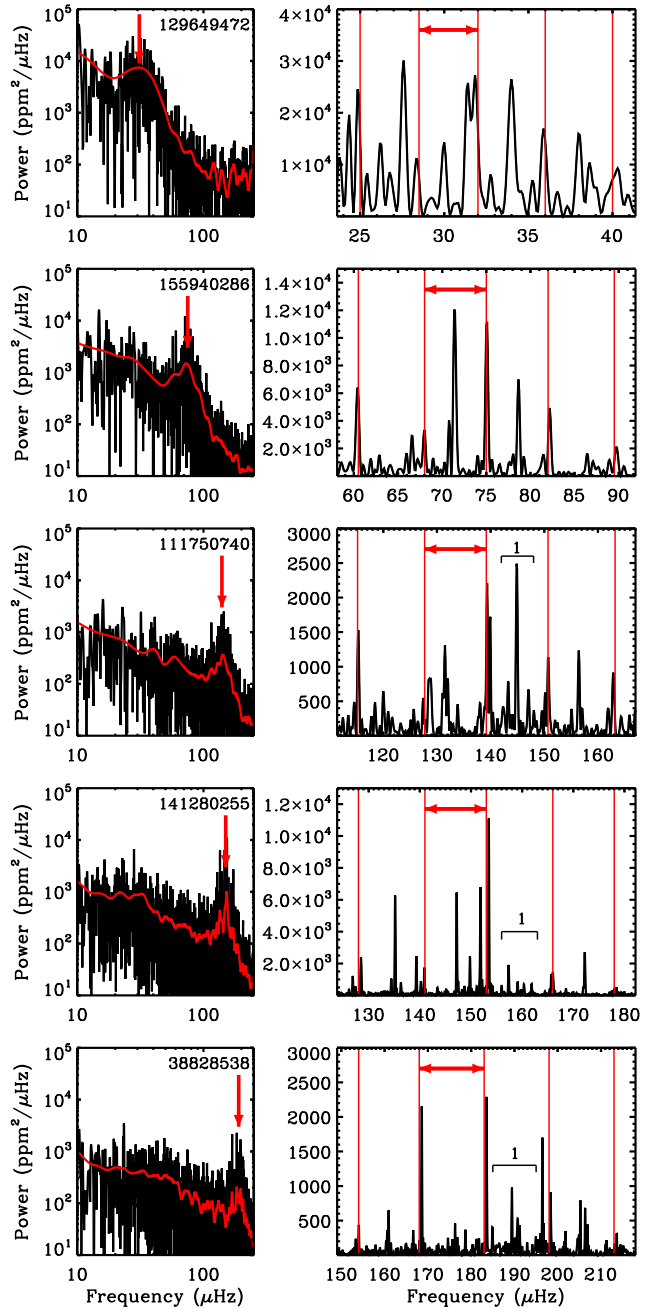}
\caption{Power spectra sample of our targets representative of the \numax\ range that they cover from around the red clump (top) to the low luminosity red giant branch (bottom). {\it Left}: Spectra shown in log-log space (smoothed in red) showing the location of the oscillation power excess, \numax, indicated by red arrows on top of a frequency-dependent granulation background and flat white noise component. {\it Right}: Close-up of spectra showing locations of the roughly equally-spaced radial modes (using red equally-spaced vertical lines to guide the eye) and their average separation, \dnu\ (red horizontal arrows). In the three bottom panels multiple dipole ($l=1$) mixed modes are resolved in between consecutive radial modes as indicated by the black brackets. \label{Fig:PS}}
\end{figure}
The stars selected were included in an early release of processed data from the TASOC pipeline\footnote{T'DA Data Release Notes - Data Release 3 for TESS Sectors 1$+$2 (\url{https://doi.org/10.5281/zenodo.2510028})}. The
calibrated full frame images were produced by the TESS Science Processing Operations Center (SPOC) at NASA Ames Research Center \citep{jenkinsSPOC2016}, and processed by combining the methodology from the K2P2 pipeline \citep{2015ApJ...806...30L} for extracting the flux from target pixel data with the KASOC filter for systematics correction \citep{2014MNRAS.445.2698H}. The resulting TASOC light curves were high-pass filtered using a filter width of 4 days, corresponding to a cut-off frequency of approximately 3\muhz\,, and $4\sigma$ outliers were removed. Finally, we used linear interpolation to fill gaps that lasted up to three consecutive cadences and derived the Fourier transforms (power frequency spectra) of each light curve.

The light curves for the seven stars observed in both sectors were merged. To follow the approach anticipated for the millions of light curves from the TESS full frame images in the future, we first applied the neural network-based detection algorithm by \citet{Hon:2018jr} resulting in detection of oscillations in the power spectra of all stars except one. The non-detection (TIC~204314449) is listed as an A2 dwarf and a 'Visual Double' in the University of Michigan Catalogue of two-dimensional spectral types for the HD stars \citep{1994ASPC...60..285H}, and hence possibly too hot to show solar-like oscillations, or potentially contaminated. For the current test case, the number of stars was small enough that we visually checked the results, which confirmed all detections and the non-detection. The power spectra of a representative sample of the stars are shown in Figure~\ref{Fig:PS} showing clear oscillation excess power and the frequency pattern required to measure both \numax\ and \dnu.

The neural network also supplies a rough estimate for \numax, which we provided as a prior to 13 independent groups analysing the power spectra to extract high-precision values of both \numax, \dnu, and their respective uncertainties using their preferred method. These methods have been thoroughly tested and described in the literature \citep[see e.g.,][]{Huber:2009ti,2009A&A...506....7G,Hekker:2010jb,Mathur:2010dl,Mosser:2011gw,Kallinger:2012gg,2014A&A...571A..71C,2016MNRAS.456.2183D,2017MNRAS.469.1360C,2019arXiv190911927Z}.

From the 13 independent determinations of the global asteroseismic parameters we adopted as central reference value for \dnu\ and \numax\ the results from the pipeline by \citet{2009A&A...506....7G}, as this method was on average closest to the ensemble mean after applying a 2-$\sigma$ outlier rejection. Uncertainties in the global asteroseismic parameters obtained by the selected pipeline are at the 1.9\% and 2.4\% level for \dnu\ and \numax\,, respectively. These uncertainties are of comparable magnitude to those obtained from a single campaign with the K2 mission \citep[see appendix in][]{2017ApJ...835...83S} and about twice as large as those extracted from 50 days of \kepler\, observations \citep[see Figs.~3~and~4 in][]{Hekker:2012ic}. We report the central values and statistical uncertainties in \dnu\ and \numax\ from the selected pipeline for all targets in Table~\ref{tab:props}.

For each star, we take into account the scatter across the different methods by adding in quadrature the standard deviation among the central values retained after the 2-$\sigma$ outlier rejection procedure to the formal uncertainty reported by the selected reference method. This consolidation process yields median uncertainties of 3.9\% in \dnu\ and 2.6\% in \numax, where the individual contribution arising from this systematic component to the total uncertainty is listed in Table~\ref{tab:props}. We note that we could decrease the level of uncertainties resulting from our `blind' statistical consolidation approach by for example checking the \dnu\ and \numax\ results against the power spectra and/or \'echelle diagrams \citep[see Fig.~5 in][]{Stello:2011hu}. However, we want to draw a realistic picture of the uncertainties one can expect when dealing with large ensembles of stars (as expected from TESS) where detailed 'boutique' analysis/checking on a star-by-star basis is not practically feasible. Hence, our quoted uncertainties are conservative, but representative for analysis of TESS red giants where several pipelines are involved.
\section{Derived stellar properties}\label{sec:stel_prop}
We have determined stellar properties for a subsample of 17 stars that had spectroscopic measurements of effective temperature and chemical composition available in the literature. Since one of our goals is to follow the same analysis procedure expected for large ensembles of stars, we assumed fixed uncertainties in \teff and [Fe/H] of 80~K and 0.08~dex, which are at the level of those provided by current large-scale spectroscopic surveys. To extract the physical properties of our sample, the atmospheric information was complemented with the asteroseismic scaling relations:
\begin{equation}\label{eqn:sca_dnu}
\left(\frac{\Delta\nu}{\Delta\nu_{\sun}}\right)^{2} \simeq 
\frac{\overline{\rho}}{\overline{\rho}_{\sun}}
\end{equation}
\begin{equation}\label{eqn:sca_num} 
\left(\frac{\nu_{\mathrm{max}}}{\nu_{\mathrm{max},\sun}}\right) \simeq
\frac{M}{M_{\sun}}
\left(\frac{R}{R_{\sun}}\right)^{-2}
\left(\frac{T_{\mathrm{eff}}}{T_{\mathrm{eff},\sun}}\right)^{1/2} \,,
\end{equation}

\noindent where we adopted $\Delta\nu_{\sun}=135.5$~($\mu$Hz) and $\nu_{\mathrm{max},\sun}=3140$~($\mu$Hz) as obtained by our reference pipeline from the analysis of solar data.

Seven teams independently applied grid-based modelling pipelines based on stellar evolution models or isochrones to determine the main physical properties of the targets \citep[see][and references therein]{Basu:2012fg,2015MNRAS.452.2127S,2017MNRAS.467.1433R,2018A&A...618A..54M,2019MNRAS.489.1753Y}. When matching the models to the atmospheric properties and the global asteroseismic parameters \dnu\, and \numax\, the pipelines yielded median uncertainties of $\sim$6\% in radius, $\sim$14\% in mass, and $\sim$50\% in age. These statistical uncertainties are of the same magnitude to those obtained with the K2 mission \citep{2019arXiv190412444S}, as expected from the similar resulting errors in the global seismic parameters described in Section~\ref{sec:data_analysis}, and about a factor of two larger than what can be achieved with the full duration of the \kepler\ observations \citep{2018ApJS..239...32P}.

In addition to the asteroseismic information, five of the pipelines can include parallaxes from Gaia DR2 \citep{GaiaCollaboration:2018dt} coupled with Tycho-2 \citep{2000A&A...355L..27H} observed $V$-magnitudes in their fitting algorithm to further constrain the stellar properties. As a consequence of having the additional constraint on stellar radius from the astrometry, the resulting uncertainties decrease to a level of $\sim$3\% in radius, $\sim$6\% in mass, and $\sim$20\% in age. This level of precision resembles that obtained with the use of the full length of asteroseismic observations from the nominal \kepler\ mission, and emphasizes the potential of TESS for Galactic studies using red giants given its larger sky coverage, simple and reproducible selection function, and one order of magnitude higher expected yield of asteroseismic detections than any other previous mission.
\begin{figure}
\includegraphics[width=\linewidth]{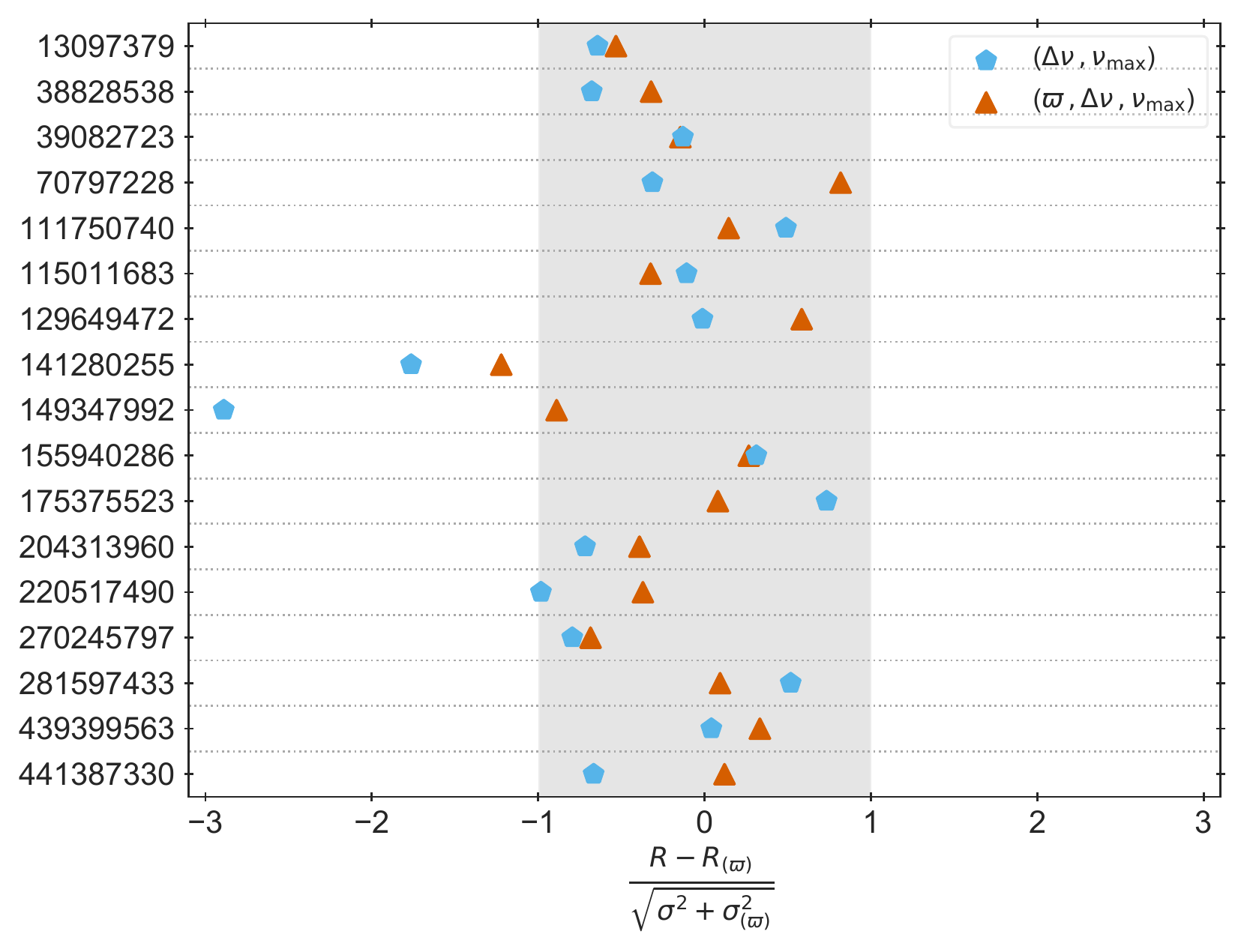}
\caption{Comparison of stellar radii obtained with {\tt BASTA} when fitting different combinations of input parameters: Gaia DR2 parallax and $V$-band magnitude ($\varpi$), global asteroseismic parameters (\dnu,\numax), and all combined ($\varpi$,\dnu,\numax). Effective temperature and composition are also fitted in all cases. See text for details.\label{Fig:BASTA_props}}
\end{figure}

To illustrate the differences in the obtained stellar properties arising from the selection of fitted observables, Fig.~\ref{Fig:BASTA_props} shows the stellar radius obtained with one of the pipelines \citep[{\tt BASTA},][]{2015MNRAS.452.2127S} when fitting different combinations of input parameters. The figure uses as the reference value the case when, in addition to the atmospheric properties, only the Gaia DR2 parallax and observed $V$-band magnitude are included in the fit. For the majority of the targets the results are consistent across the three sets within their formal statistical uncertainties. A summary of the measured and derived stellar properties for our targets can be found in Table~\ref{tab:props}, where we have listed the central values and statistical uncertainties obtained with the {\tt BASTA} pipeline, and determined the systematic contribution as the standard deviation across the results reported by all pipelines.

Two targets (TIC~141280255 and TIC~149347992) present a larger disagreement between the radii obtained with parallax and the seismic set (\dnu, \numax). We investigated if these discrepancies were due to the quality of the astrometric data by computing the re-normalised unit weight error (RUWE\footnote{see Gaia technical note  GAIA-C3-TN-LU-LL-124-01 (\url{https://www.cosmos.esa.int/web/gaia/dr2-known-issues})}) for our sample of stars. In the case of TIC~141280255 we obtained a RUWE=1.98, which is above the value recommended by the Gaia team as a criterion for a good astrometric solution (RUWE$\leq1.4$). Therefore, we adopt for this star the stellar properties obtained from fitting the asteroseismic input only (\dnu, \numax).

In the case of TIC~149347992 the discrepancy is the result of predicted evolutionary phases: while the parallax-only solution suggests that the star in the clump phase, the asteroseismic fit favours a star in the red-giant branch. The combined fit therefore presents a bimodal distribution that encompasses these two families of solutions. A similar situation occurs in the fit of TIC~175375523, which shows agreement in the radius determined from different sets of input but has a fractional age uncertainty above unity when only (\dnu, \numax) are included in the fit. Its resulting age distribution is bimodal in this set as both red-giant branch and clump models can reproduce the observations, but the inclusion of parallax information favours the red giant branch solution and accounts for the $\sim17$\% statistical uncertainty reported in Table~\ref{tab:props}. The availability of evolutionary classifications from deep neural networks trained on short \kepler\ data \citep{Hon:2018jr} would further decrease the obtained uncertainties by clearly disentangling these two scenarios.
\begin{figure}
\includegraphics[width=\linewidth]{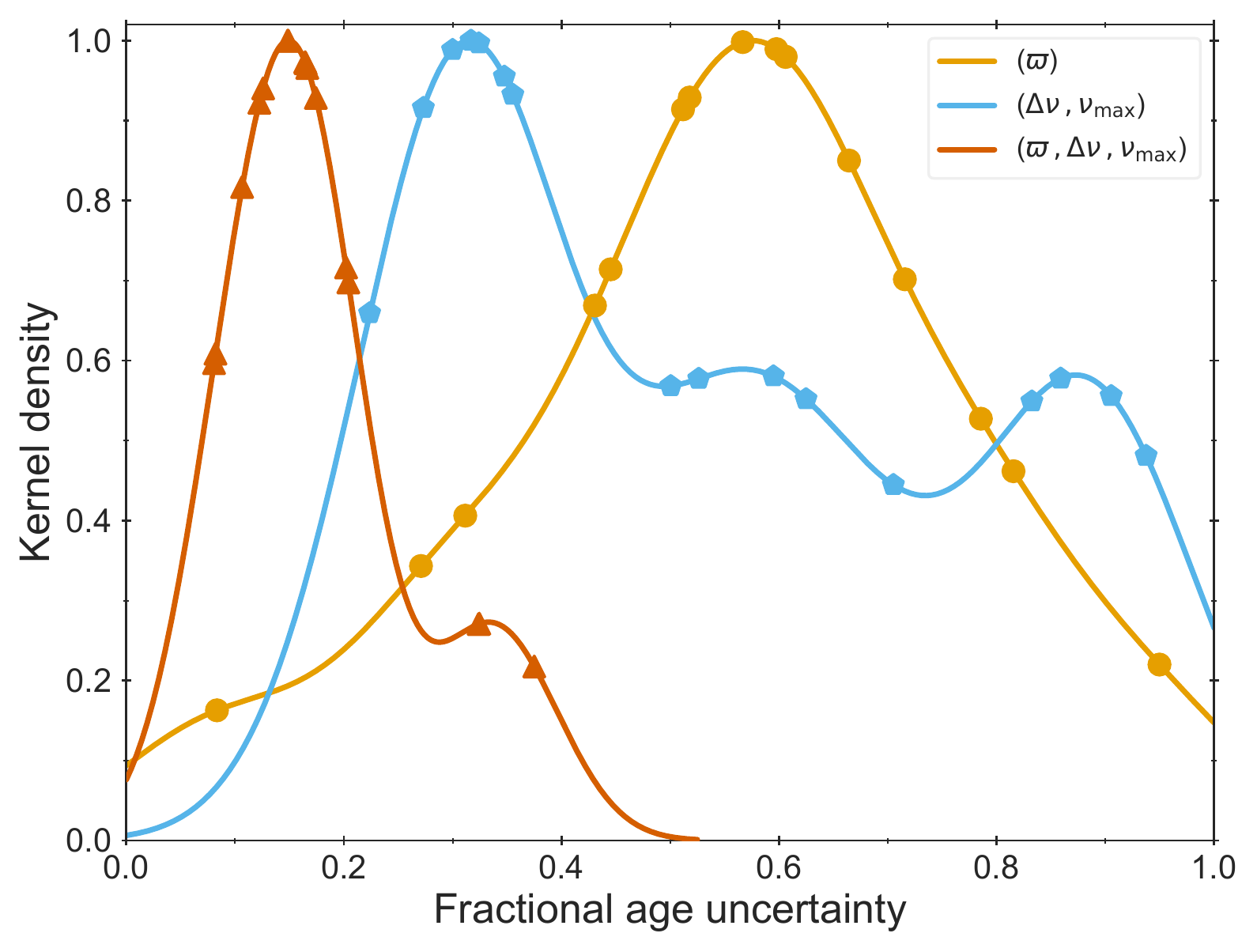}
\caption{Distribution of fractional age uncertainties for our sample of stars determined by the {\tt BASTA} pipeline fitting different combinations of available observables. The points indicate the individual values used to construct the Gaussian kernel density estimation. For better visualization we have excluded TIC~175375523 from the figure. See text for details.\label{Fig:BASTA_unc}}
\end{figure}

In Fig.~\ref{Fig:BASTA_unc} we plot the distribution of fractional age uncertainties obtained with {\tt BASTA} for the three considered cases of input, showing the clear improvement in precision when asteroseismic information and parallax are simultaneously included in the fit. For visualization purposes we have excluded the target TIC~175375523 from the figure. Our stellar ages at the 20\% level are significantly more precise than what is obtained by data-driven and neural-network methods trained using asteroseismic ages from \kepler\ \citep[above the 30\% level, see e.g.,][]{2019MNRAS.489..176M}. As a final remark, we note that asteroseismically derived properties of red giants are accurate to at least a similar level than our statistical uncertainties \citep[below $\sim5$\% and $\sim10$\% for radii and masses, respectively. See discussion in e.g.,][and references therein]{2018ApJS..239...32P}. We have made emphasis on our achieved precision instead of accuracy as our results could still be affected by a systematic component arising from uncertainties in evolutionary calculations, although recent investigations quantifying these effects at solar metallicity suggest that they are smaller than our statistical uncertainties \citep{2019arXiv191204909S}.
\section{Conclusions}\label{sec:conc}
We presented the first ensemble analysis of red giants stars observed with the TESS mission. We selected a sample of 25 stars where we expected to detect oscillations based on their magnitude and parallax value, and analysed the extracted light curves in search for asteroseismic signatures in the power spectra. Our main findings can be summarized as follows:
\begin{itemize}
\item We detected oscillations in all the stars (except one that was likely incorrectly listed as a red giant). Despite the modest number of stars in our sample, our detection yield supports that the TESS photometric performance is similar to that of \kepler\ and K2 except shifted by about 5 magnitudes towards brighter stars due to its smaller aperture.
\item Individual pipelines retrieve the global asteroseismic parameters with uncertainties at the $\sim$2\% level in \dnu\ and $\sim$2.5\% in \numax\,, which respectively increase to $\sim$4\% and $\sim$3.5\% when we take into account the scatter across results. We consider these uncertainties to be representative for the forthcoming ensemble analysis of TESS targets observed in 1-2 sectors, as individual validation of the results will not be feasible due to the large number of targets observed.
\item Grid-based modelling techniques applying asteroseismic scaling relations were used to retrieve stellar properties for the 17 targets with spectroscopic information. Radii, masses, and ages were obtained with uncertainties at the 6\%, 14\%, and 50\% level, and decrease to 3\%, 6\%, and 20\% when parallax information from Gaia DR2 is included.
\end{itemize}

The expected number of red giants with detected oscillations by TESS ($\sim$500,000\footnote{Based on a preliminary simulation of the full TESS sky (TESS GI Proposal No G011188).}) greatly surpasses the final yield of \kepler\ ($\sim$20,000). In this respect, the combination of TESS observations, Gaia astrometry, and large scale spectroscopic surveys holds a great potential for studies of Galactic structure where precise stellar properties (particularly ages) are of key importance. We note that the recently approved extended TESS mission will change the 30-min cadence to 10 minutes, making it possible to detect oscillations of stars of smaller radii using the full frame images. This will enable more rigorous investigations of the asteroseismic mass scale for giants when anchored to empirical mass determinations (e.g., from eclipsing binaries) of turn-off and subgiant stars.
\bigskip
\acknowledgments
This paper includes data collected by the TESS mission, which are publicly available from the Mikulski Archive for Space Telescopes (MAST). Funding for the TESS mission is provided by NASA’s Science Mission directorate. Funding for the TESS Asteroseismic Science Operations Centre is provided by the Danish National Research Foundation (Grant agreement no.: DNRF106), ESA PRODEX (PEA 4000119301) and Stellar Astrophysics Centre (SAC) at Aarhus University. VSA acknowledges support from the Independent Research Fund Denmark (Research grant 7027-00096B). DB is supported in the form of work contract FCT/MCTES through national funds and by FEDER through COMPETE2020 in connection to these grants: UID/FIS/04434/2019; PTDC/FIS-AST/30389/2017 \& POCI-01-0145-FEDER-030389. LB, RAG and BM acknowledge the support from the CNES/PLATO grant. DB acknowledges NASA grant NNX16AB76G. TLC acknowledges support from the European Union's Horizon 2020 research and innovation programme under the Marie Sk\l{}odowska-Curie grant agreement No.~792848 (PULSATION). This work was supported by FCT/MCTES through national funds (UID/FIS/04434/2019). EC is funded by the European Union’s Horizon 2020 research and innovation program under the Marie Sklodowska-Curie grant agreement No. 664931. RH and MNL acknowledge the support of the ESA PRODEX programme. T.S.R acknowledges financial support from Premiale 2015 MITiC (PI B. Garilli). KJB is supported by the National Science Foundation under Award AST-1903828. MSL is supported by the Carlsberg Foundation (Grant agreement no.: CF17-0760). MC is funded by FCT//MCTES through national funds and by FEDER through COMPETE2020 through these grants: UID/FIS/04434/2019, PTDC/FIS-AST/30389/2017 \& POCI-01-0145-FEDER-030389, CEECIND/02619/2017. The research leading to the presented results has received funding from the European Research Council under the European Community's Seventh Framework Programme (FP7/2007-2013) / ERC grant agreement no 338251 (StellarAges). AM acknowledges support from the European Research Council Consolidator Grant funding scheme (project ASTEROCHRONOMETRY, G.A. n. 772293, http://www. asterochronometry.eu). AMS is partially supported by MINECO grant ESP2017-82674-R. JCS acknowledges funding support from Spanish public funds for research under projects ESP2017-87676-2-2, and from project RYC-2012-09913 under the ’Ramón y Cajal’ program of the Spanish Ministry of Science and Education. Resources supporting this work were provided by the NASA High-End Computing (HEC) Program through the NASA Advanced Supercomputing (NAS) Division at Ames Research Center for the production of the SPOC data products.
\bibliography{TESSwg7}

%
\begin{splitdeluxetable*}{lllllllBlllll}
\tablecaption{Measured and derived stellar properties of our targets. Observed $V$-magnitudes are extracted from the Tycho-2 catalogue. The global asteroseismic quantities and stellar properties include a statistical and systematic component derived as described in Sections~\ref{sec:data_analysis} and~\ref{sec:stel_prop}, respectively. We report them here as ${\it value}\pm\sigma_\mathrm{\it sta}\pm\sigma_{\it sys}$.\label{tab:props}}
\tabletypesize{\scriptsize}
\tablehead{
\colhead{TIC} & \colhead{HIP} & \colhead{\numax} & \colhead{\dnu} & \colhead{$V$} & \colhead{\teff} &
\colhead{[Fe/H]} & \colhead{TIC} &\colhead{R} & \colhead{M} & \colhead{Age} & \colhead{Atmospheric Properties}\\
\nocolhead{} & \nocolhead{} & \colhead{($\mu$Hz)} & \colhead{($\mu$Hz)} & \colhead{(mag)} & \colhead{(K)} &\colhead{(dex)} & \nocolhead{} & \colhead{(R$_\sun$)} & \colhead{(M$_\sun$)} & \colhead{(Gyr)} & \nocolhead{}
}
\startdata
13097379&114842&$59.10 \pm1.50\pm1.01$&$6.02 \pm0.03\pm0.24$&$6.646\pm0.010$&$4634\pm80 $&$0.04 \pm0.08$&13097379&$8.49 \pm0.28\pm0.17$&$1.22\pm0.08\pm0.06$&$6.10 \pm1.06\pm0.97$& \text{\citet{2015AJ....150...88L}}\\
38574220&19805&$29.40 \pm0.90\pm0.72$&$4.06 \pm0.20\pm0.26$&$5.577\pm0.009$&--&--&38574220&--&--&--&--\\
38828538&21253&$189.90\pm1.60\pm0.42$&$14.90\pm0.10\pm0.13$&$5.896\pm0.009$&$4828\pm80 $&$0.11 \pm0.08$&38828538&$4.66 \pm0.10\pm0.06$&$1.21\pm0.05\pm0.03$&$6.20 \pm0.50\pm1.02$& \text{\citet{2015MNRAS.448.2749A}}\\
39082723&4293&$49.30 \pm2.10\pm1.99$&$5.20 \pm0.10\pm0.06$&$5.574\pm0.009$&$4706\pm80 $&$-0.05\pm0.08$&39082723&$9.30 \pm0.27\pm0.17$&$1.19\pm0.09\pm0.07$&$5.90 \pm1.20\pm1.37$& \text{\citet{2015MNRAS.448.2749A}}\\
47424090&112612&$28.30 \pm1.80\pm1.90$&$3.40 \pm0.10\pm0.29$&$6.930\pm0.010$&--&--&47424090&--&--&--&--\\
70797228&655&$31.80 \pm1.50\pm0.75$&$4.37 \pm0.20\pm0.37$&$5.787\pm0.009$&$4750\pm80 $&$0.12 \pm0.08$&70797228&$11.27\pm0.61\pm0.47$&$1.19\pm0.13\pm0.11$&$6.80 \pm2.20\pm2.20$& \text{\citet{2011AnA...536A..71J}}\\
77116701&103071&$48.30 \pm7.60\pm29.85$&$5.64 \pm0.20\pm3.37$&$8.568\pm0.018$&--&--&77116701&--&--&--&--\\
111750740&113148&$142.60\pm2.70\pm1.11$&$11.80\pm0.10\pm0.23$&$5.658\pm0.009$&$4688\pm80 $&$0.16 \pm0.08$&111750740&$5.11 \pm0.16\pm0.09$&$1.06\pm0.07\pm0.05$&$10.80\pm1.78\pm1.71$& \text{\citet{2016AJ....152...19W}}\\
115011683&103836&$58.80 \pm1.20\pm0.97$&$6.10 \pm0.10\pm0.01$&$6.057\pm0.010$&$4590\pm80 $&$-0.13\pm0.08$&115011683&$7.92 \pm0.21\pm0.22$&$1.04\pm0.07\pm0.07$&$9.90 \pm1.63\pm2.01$& \text{\citet{2016AJ....152...19W}}\\
129649472&105854&$31.80 \pm1.20\pm1.39$&$4.20 \pm0.20\pm0.11$&$5.755\pm0.009$&$4748\pm80 $&$0.28 \pm0.08$&129649472&$10.85\pm0.60\pm0.24$&$1.13\pm0.12\pm0.06$&$8.50 \pm2.76\pm1.87$& \text{\citet{2015AnA...574A..50J}}\\
139756492&106566&$27.60 \pm0.90\pm0.35$&$4.16 \pm0.20\pm0.65$&$6.819\pm0.010$&--&--&139756492&--&--&--&--\\
141280255&25918&$150.40\pm1.00\pm0.57$&$12.52\pm0.02\pm0.10$&$5.307\pm0.009$&$4630\pm80 $&$0.33 \pm0.08$&141280255&$4.98 \pm0.12\pm0.05$&$1.07\pm0.06\pm0.03$&$11.70\pm2.62\pm2.39$& \text{\citet{2008AnA...484L..21M}}\\
144335025&117075&$68.50 \pm1.60\pm0.64$&$7.35 \pm0.20\pm0.39$&$6.194\pm0.010$&--&--&144335025&--&--&--&--\\
149347992&26190&$165.80\pm4.10\pm16.12$&$11.10\pm0.40\pm0.60$&$6.405\pm0.010$&$5132\pm80 $&$-0.17\pm0.08$&149347992&$7.20 \pm0.38\pm0.20$&$2.17\pm0.22\pm0.06$&$0.80 \pm0.30\pm0.10$& \text{\citet{1999AnA...348..487R}}\\
155940286&1766&$73.20 \pm1.30\pm0.32$&$7.40 \pm0.02\pm0.11$&$6.810\pm0.010$&$4630\pm80 $&$0.03 \pm0.08$&155940286&$6.95 \pm0.18\pm0.14$&$1.01\pm0.06\pm0.04$&$12.00\pm1.78\pm36.07$& \text{\citet{2016AJ....152...19W}}\\
175375523&114775&$60.00 \pm1.10\pm0.30$&$5.80 \pm0.10\pm0.14$&$5.899\pm0.009$&$4660\pm80 $&$0.26 \pm0.08$&175375523&$9.00 \pm0.30\pm0.58$&$1.38\pm0.09\pm0.21$&$4.50 \pm0.76\pm1.35$& \text{\citet{2011AnA...536A..71J}}\\
183537408&117659&$57.90 \pm1.10\pm0.80$&$6.20 \pm0.20\pm0.42$&$6.781\pm0.010$&--&--&183537408&--&--&--&--\\
204313960&113801&$106.00\pm3.30\pm1.47$&$9.40 \pm0.50\pm0.43$&$6.083\pm0.010$&$4897\pm80 $&$-0.20\pm0.08$&204313960&$6.50 \pm0.18\pm0.13$&$1.32\pm0.07\pm0.05$&$3.80 \pm0.57\pm0.57$& \text{\citet{1999AnA...348..487R}}\\
220517490&12871&$117.30\pm1.20\pm0.60$&$10.87\pm0.02\pm0.15$&$5.846\pm0.009$&$4961\pm80 $&$-0.26\pm0.08$&220517490&$5.61 \pm0.11\pm0.10$&$1.10\pm0.04\pm0.04$&$6.10 \pm0.50\pm1.18$& \text{\citet{2015MNRAS.448.2749A}}\\
237914586&17440&$47.00 \pm1.40\pm1.33$&$5.74 \pm0.20\pm0.08$&$3.959\pm0.009$&--&--&237914586&--&--&--&--\\
270245797&109584&$72.20 \pm1.70\pm0.64$&$6.60 \pm0.10\pm0.27$&$6.239\pm0.009$&$4824\pm80 $&$-0.10\pm0.08$&270245797&$8.73 \pm0.23\pm0.39$&$1.60\pm0.09\pm0.13$&$2.10 \pm0.22\pm0.64$& \text{\citet{2015MNRAS.448.2749A}}\\
281597433&2789&$73.30 \pm1.00\pm1.37$&$7.20 \pm0.03\pm0.19$&$6.163\pm0.010$&$4700\pm80 $&$-0.41\pm0.08$&281597433&$6.77 \pm0.15\pm0.24$&$0.95\pm0.05\pm0.08$&$11.00\pm1.35\pm2.73$& \text{\citet{1999AnA...348..487R}}\\
439399563&343&$44.30 \pm1.40\pm0.66$&$4.54 \pm0.06\pm0.11$&$5.892\pm0.009$&$4778\pm80 $&$0.11 \pm0.08$&439399563&$10.69\pm0.34\pm0.40$&$1.44\pm0.11\pm0.14$&$3.50 \pm0.71\pm0.78$& \text{\citet{2015AnA...580A..24D}}\\
441387330&102014&$46.60 \pm0.80\pm0.86$&$5.25 \pm0.10\pm0.25$&$5.592\pm0.009$&$4710\pm80 $&$-0.02\pm0.08$&441387330&$10.18\pm0.34\pm0.68$&$1.40\pm0.09\pm0.20$&$3.40 \pm0.57\pm1.19$& \text{\citet{2011AnA...536A..71J}}\\
\enddata
\tablecomments{Last column gives the reference from which we retrieved the central values of \teff\ and [Fe/H] used for the grid-based modelling. Their uncertainties have been homogenised to 80~K and 0.08~dex, respectively (see Section~\ref{sec:stel_prop}).}
\end{splitdeluxetable*}

\end{document}